\documentclass[10pt,a4paper]{article}
\usepackage{fullpage}
\usepackage{amsmath,amssymb,graphicx,paralist}
\usepackage[round]{natbib}
\include{psfig}
\begin{document}

\title{
Comment on ``Bayesian astrostatistics: a backward look to the future" by Tom
Loredo, arXiv:1208.3036}

\author{S. Andreon\\
INAF-Osservatorio Astronomico di Brera, via Brera 28, 20121, Milano, Italy
}

\maketitle

\begin{abstract}
This short note points out two of the incongruences that I find in
the \citet{loredo12} comments on \citet{andreon2012bayesian}, i.e. on
my chapter
written for the book  ``Astrostatistical Challenges for the New Astronomy". 
First, I find illogic the 
Loredo decision of putting my chapter among those presenting simple
models, because one of the models illustrated in my chapter is 
qualified by him as ``impressing for his complexity".
Second, Loredo criticizes my chapter at one location confusing it with another
paper by another author, because my chapter do not touch the subject
mentioned by \citet{loredo12} critics, the comparison between Bayesian
and frequentist fitting models.
\end{abstract}

\citet{loredo12} paper, ``Bayesian astrostatistics: a backward look to
the  future" is, as indicated in his Comment line, a lightly revised
version  of a chapter in ``Astrostatistical Challenges for the New
Astronomy",  the inaugural volume for the Springer Series in
Astrostatistics. This volume also includes a chapter written
by me \citep{andreon2012bayesian}, and one by March et al. 
My chapter aims to show the 
simplicity of performing a Bayesian statistical analysis
with two examples, the first one is original
and has been first presented in \citet{andreon2010scaling}, 
the second example is drawn from \citet{March11}. In Sect. 4.3 of
my chapter I also consider a model with increased complexity 
by inserting new nodes in the \citet{March11} model
to check model adequacy.

I disagree with several \citet{loredo12} comments on my
chapter, but I will restrict my attention here to 
two of them.

\citet{loredo12} paper\footnote{At the time of this writing 
there is one single version of the paper, arXiv:1208.3036v1, and this is the
version which I'm referring to.}
splits chapters of the volume in two groups according to the complexity
of the used models. 
The first one is composed by chapters using
simpler models and my chapter belongs to this group according
to \citet{loredo12}.
The other group is composed by contributions using
more complex models, and includes the
March et al. chapter.
The  March et al. model
is commented by him as ``impressing for his complexity" (his pag 20).

Unfortunately for \citet{loredo12}, the March et al. model
is also the model adopted in my second example, and thus  my chapter
unambigously deals with complex models, and must be put among
those dealing with complex models.
To be pedant, in Sect. 4.3 of my chapter I add complexity to
March et al. model, and in Sect. 3 I present one more model, developed by 
myself \& Hurn, of similar complexity.

About the second incongruence,
\citet{loredo12} paper, after referring a few times to my chapter as
``Andreon's" chapter, topic, contribution or just ``Andreon's", comments:
``In the context of nonlinear regression with measurement error
--Andreon's topic-- Carroll et al. (2006) provides a useful entry
point to both Bayesian and frequentist literature, incidentally also
describing a number of frequentist approaches to such problems that
would be more fair competitors to Bayesian MLMs than the $\chi^2$
approach that serves as Andreon's straw man competitor." 

Unfortunately for \citet{loredo12}, in my chapter \citep{andreon2012bayesian}
there is no 
comparison between 
a Bayesian and $\chi^2$ fit or any other frequentist fitting method. 
There are no frequentist approaches to regression in my chapter.
There is no competition at all between methods, because only a single
method (Bayesian) is used to fit the data. \citet{loredo12} is
criticizing a comparison between frequentist and
Bayesian fitting models, but there is
no trace of it in my chapter, simply and plainly he is confusing
my chapter with another paper,
although my chapter get his criticisms.

\medskip
To sum up, 
I find illogic the 
\citet{loredo12} decision of putting my chapter among those 
presenting simple
models, given his claim that one of models illustrated in my chapter
is ``impressing for his complexity" and the other one has similar complexity.
Second, \citet{loredo12} criticizes my chapter at one location
confusing it with another
paper by another author, because my chapter do not touch the subject
criticized by Loredo, the comparison of frequentist and
Bayesian fitting models.

\bibliography{book}

\begin{thebibliography}{4}
\providecommand{\natexlab}[1]{#1}
\providecommand{\url}[1]{\texttt{#1}}
\expandafter\ifx\csname urlstyle\endcsname\relax
  \providecommand{\doi}[1]{doi: #1}\else
  \providecommand{\doi}{doi: \begingroup \urlstyle{rm}\Url}\fi

\bibitem[Andreon(2012)]{andreon2012bayesian}
S.~Andreon.
\newblock {Understanding better (some) astronomical data using Bayesian
  methods}.
\newblock \emph{Astrostatistical Challenges for the New Astronomy, edited by J.
  Hilbe.}, Publisher: Springer Series on Astrostatistics, in press,
  arXiv:1112.3652, 2012.

\bibitem[Andreon and Hurn(2010)]{andreon2010scaling}
S.~Andreon and M.A. Hurn.
\newblock {The scaling relation between richness and mass of galaxy clusters: a
  Bayesian approach}.
\newblock \emph{Monthly Notices of the Royal Astronomical Society},
  404\penalty0 (4):\penalty0 1922--1937, 2010.
\newblock ISSN 1365-2966.

\bibitem[{Loredo}(2012)]{loredo12}
T.~J. {Loredo}.
\newblock {Bayesian astrostatistics: a backward look to the future}.
\newblock \emph{Astrostatistical Challenges for the New Astronomy, edited by J.
  Hilbe.}, Publisher: Springer Series on Astrostatistics, in press,
  arXiv:1208.3036, 2012.

\bibitem[{March} et~al.(2011){March}, {Trotta}, {Berkes}, {Starkman}, and
  {Vaudrevange}]{March11}
M.~C. {March}, R.~{Trotta}, P.~{Berkes}, G.~D. {Starkman}, and P.~M.
  {Vaudrevange}.
\newblock {Improved constraints on cosmological parameters from Type Ia
  supernova data}.
\newblock \emph{Monthly Notices of the Royal Astronomical Society},
  418:\penalty0 2308--2329, 2011.

\end{thebibliography}
\bibliographystyle{plainnat}

\end{document}